\documentclass[12pt]{article}
\usepackage{amsmath,amsfonts,amssymb,amsthm,mathtools,mathrsfs,tikz}
\usepackage{scalerel}			
\usepackage{fontawesome}		
\usepackage{mathabx}			
\usepackage{bbm}				
\usepackage{authblk}			
\usepackage[letterpaper, top=1in, bottom=1in, left=.75in, right=.75in]{geometry}
\usepackage{setspace}
\usepackage[pdfencoding=auto,psdextra,colorlinks=TRUE,linkcolor=black,citecolor=black,urlcolor=blue]{hyperref}
\usepackage{bookmark}			
\newfont{\boldit}{cmbxti12}
\newcommand{\C}{\mathbb{C}}  				
\newcommand{\R}{\mathbbm{R}}  				
\newcommand{\N}{\mathbbm{N}}  				
\newcommand{\sgn}{\text{sgn}}					

\newcommand{\Ckd}[2]{C^{#1}(\R^{#2})}		
\newcommand{\DTd}[1]{\mathcal{D}(\R^{#1})}	
\newcommand{\DDd}[1]{\mathcal{D}\,'(\R^{#1})}	
\newcommand{\Lpd}[2]{L^{#1}(\R^{#2})}		
\newcommand{\fall}{\;\;\forall\;}			

\renewcommand{\geq}{\vargeq}

\makeatletter
\def\thmheadbrackets#1#2#3{%
  \thmname{#1}\thmnumber{\@ifnotempty{#1}{ }\@upn{#2}}%
  \thmnote{ {\the\thm@notefont[\,#3\,]}}}
\makeatother

\newtheoremstyle{brakets}
  {1.5em}
  {1.5em}
  {\itshape}
  {}
  {\bfseries}
  {\\}
  { }
  {\thmheadbrackets{#1}{#2}{#3}}

\theoremstyle{brakets}
\newtheorem{thm}{Theorem}[section]
\newtheorem{cor}[thm]{Corollary}
\newtheorem{defn}[thm]{Definition}

\newtheorem{lem}[thm]{Lemma}
\newtheorem{prop}[thm]{Proposition}

\numberwithin{equation}{section}

\title{The Energy Eigenvalue for the Singular Wave Function of the Three Dimensional Dirac Delta Schrodinger Potential via Distributionally Generalized Quantum Mechanics}
\author{Dr. Michael Maroun, PhD\footnote{Vice President of Research \& Development, TDI Corporate Labs, Austin, TX \url{TeXDynIndustries.com}}}
\affil{TeXDyn Industries Corporate Laboratories \\
Austin, TX}
\date{\today}

\begin{document}
\maketitle

\begin{abstract}
Unlike the situation for the 1d Dirac delta derivative Schrodinger pseudo potential (SPP) and the 2d Dirac delta SPP, where the indeterminacy originates from a lack of scale in the first and both a lack of scale as well as the wave function not being well defined at the support of the generalized function SPP; the obstruction in 3d Euclidean space for the Schrodinger equation with the Dirac delta as a SPP only comes from the wave function (the $L^2$ bound sate solution) being singular at the compact point support of the Dirac delta function (measure). The problem is solved here in a completely mathematically rigorous manner with no recourse to renormalization nor regularization. The method involves a distributionally generalized version of the Schrodinger theory as developed by the author, which regards the formal symbol "$H\psi$" as an element of the space of distributions, the topological dual vector space to the space of smooth functions with compact support. Two main facts come to light. The first is the bound state energy of such a system can be calculated in a well-posed context, the value of which agrees with both the mathematical and theoretical physics literature. The second is that there is then a rigorous distributional version of the Hellmann-Feynman theorem.
\end{abstract}

\section{Introduction}

The problem from the outset is that when the potential in the usual Schrodinger equation is a tempered distribution, even as mild as the Dirac measure without any derivatives (since such derivatives ruin the measure property as no such Radon-Nikodym derivative exists for measures that are not absolutely continuous with respect to Lebesgue measure), the Hamiltonian fails to make sense as a linear operator because the formal symbol asks for the sum of the Laplace operator (a linear operator) with that of a linear functional, the tempered distribution. One can begin to alleviate this problem in the following way.

Let $\DTd{d}$ be the space of infinitely differentiable functions with compact support, called test functions, and let $\DDd{d}$ be the topological dual vector space. One can now regard the formal symbol $H\psi$ as an element of $\DDd{d}$. The exact conditions for $(H\psi)\in\DDd{d}$ are sufficiently varied that it extends many cases of interest. The present problem is one such example of the direct application of an useful extension. Nonetheless, there are still examples that are of interest for which $(H\psi)$ is not a distribution. This partially occurs in the present problem where the definition for a rather singular distribution on $\R^3$ needs elucidating, while in $\R^2$ the "product" $\delta\psi$ does not correspond to any distribution. Thus, the issue for generalization stems from the product of the potential with the wave function. The kinetic term is easily replaced with the distributional Laplace operator. But the product of distributions is notoriously not robust enough for arbitrary applications, nor is the convolution\cite{AlG, Shu}.

As an aside regarding motivation, there are two very poignant reasons for extending $(H\psi)$ into the space of distributions. Firstly in the theory of scattering, the wave function is not an element of the usual Hilbert space of square integrable functions, and so to treat both scattering theory and bound state theory on equal footing one must have a well-defined notion of the Schrodinger dynamics when it comes to the Gelfand triple. This is explained more in the following paragraph that justifies the theory's name. Secondly, there is a definition of the locally square integrable function space, which necessarily involves a special class of distributions. This allows one to port the Max Born probability interpretation into the Gelfand triple, insofar as the special class of distributions is concerned. Physically, this is embodied in the fact that arguably every spatial component of any experiment ever has been conducted on a compact subset of the three Euclidean submanifold of the hyperbolic space-time manifold. The mathematical facts are embodied in the next 2 definitions, and is followed by a lemma that proves the two definitions are in fact equivalent.
\begin{defn}[Locally Square Integrable Functions v1] \label{LSIFv1}

Let $T:\R^d\to\C$, then one says $T\in L^2_{\text{loc}}(\R^d)$ if it is the case that
\[
\int\limits_{x\,\in K\subset\R^d} |T|^2\;\text{d}x < \infty \quad\fall K\subset\R^d.
\] 
\end{defn}

\begin{defn}[Locally Square Integrable Functions v2] 
\label{LSIFv2} 
\[
 L^2_{\text{loc}}(\R^d) := \left\{\; T\in\DDd{d}\,:\forall\;\text{\raisebox{2pt}{$\varphi$}}\in\DTd{d},\;\text{one has}\;\, (\text{\raisebox{2pt}{$\varphi$}} \,T)\in\Lpd{2}{d} \;\right\}
\]
\end{defn}

\begin{lem}[Bourbaki-Strichartz Equivalence] \label{BSE}
\phantom{-}
\begin{center}
{\bf {\it Definition\;\ref{LSIFv1}}} $\Longleftrightarrow$ {\bf {\it Definition\;\ref{LSIFv2}}}
\end{center}
\phantom{-}
\end{lem}
\begin{proof}[Proof of Lemma \ref{BSE} (Bourbaki-Strichartz Equivalence)]
First, definition  \ref{LSIFv1} implies definition \ref{LSIFv2}.
	\[
	\|\text{\raisebox{2pt}{$\varphi$}}\|^2_{\text{\raisebox{-2pt}{$\infty$}}}\!\!\!\int\limits_{x\,\in K} |T|^2\;\text{d}x \geq
	\int\limits_{x\,\in K} |\text{\raisebox{2pt}{$\varphi$}}|^2\,|T|^2\;\text{d}x = \int\limits_{x\,\in U} |\text{\raisebox{2pt}{$\varphi$}}\,T|^2\;\text{d}x
	\]
Now, it is shown that definition \ref{LSIFv2} implies definition \ref{LSIFv1}. Let $\R\ni d := d(K,\,\partial U) > 0$, where the notation $d(\mathbbm{A},\,\partial\mathbbm{B})$ stands for the family of metric distances from points inside the set $\mathbbm{A}$ to points on the boundary of the set $\mathbbm{B}$. Also, let $\varepsilon$ be such that $d > 2\varepsilon > 0$. Now consider the compact sets $K_{\varepsilon}$ and $K_{2\varepsilon}$ as closed $(\varepsilon,\,2\varepsilon)$-neighborhoods of $K$. One  has the set inclusion $K\subset K_{\varepsilon}\subset K_{2\varepsilon}\subset U$ with $d_{K_\varepsilon} - \varepsilon > \varepsilon > 0$.	Now define a mollifier, \raisebox{2pt}{$\varphi$}$_\varepsilon$, and take $\text{\raisebox{2pt}{$\chi$}}_{_{K_\varepsilon}}$ as the characteristic function supported on the compact set $K_\varepsilon$. The convolution $\text{\raisebox{2pt}{$\varphi$}}_{\varepsilon}\ast \text{\raisebox{2pt}{$\chi$}}_{_{K_\varepsilon}}=:\text{\raisebox{2pt}{$\varphi$}}_{_K}$, together with its inequalities lead to $|\text{\raisebox{2pt}{$\varphi$}}_{_K}|^2 \geq
	|\text{\raisebox{2pt}{$\chi$}}_{_{K}}|^2$. Hence, one sees
	\[
	\infty > \int\limits_{x\,\in U} |T|^2\, |\text{\raisebox{2pt}{$\varphi$}}_{_K}|^2\;\text{d}x \geq
	\int\limits_{x\,\in U} |T|^2\, |\text{\raisebox{2pt}{$\chi$}}_{_K}|^2\;\text{d}x = \int\limits_{x\,\in K} |T|^2\;\text{d}x. \qedhere
	\]
\end{proof}

Now let the term "distributionally generalized" be justified.
\begin{defn}[C-Spectrum]\label{csp}
Let $\langle\cdot,\,\cdot\rangle$ denote the Schwartz bracket linear functional that pairs a test function $\varphi\in\DTd{d}$ with a distribution $T\in\DDd{d}$ as,
\[
\langle T,\,\varphi\rangle\,= T(\varphi) \in\R\subseteq\C.
\]
Then whenever $\langle H\psi,\,\varphi\rangle\,= E\langle\psi,\,\varphi\rangle$, $E$ is said to be in the distributionally generalized spectrum called the \textbf{$\mathbf{ C}$-spectrum} \footnote{The term "generalized eigenvalue" already has a precise mathematical definition and is of great practical importance.}\footnote{Here the letter $C$ is not the third letter of the Latin alphabet but rather the nineteenth letter of the Russian alphabet in honor of Sergei Sobolev.}. 
\end{defn}

This definition is rather useful because it happens that there are an infinity of linear operators on infinite dimensional vector spaces including the Hilbert space of square integrable functions which are not self-adjoint, nor even normal but have real eigenvalues. There is an interesting theory that also generalizes quantum theory called $\mathcal{PT}$-symmetric quantum theory put forth by Carl Bender, which addresses this issue as well \cite{Ben}. In addition, notice that the definition \ref{csp} includes the possibility of generalized eigenvalues since $\psi$ need not be in $\Lpd{2}{d}$.

It may happen that the relation does not hold for all test functions but for some non-trivial\footnote{By non-trivial, it is meant that the set is neither empty nor does it have zero as its only element. It may be the case that there is an unique test function that makes the expression valid, or that there is an unique function but that it is not a test function, which is needed. In this last case, see the notion of proxy test function that follows.} subset of them. Even still, it is also possibly the case that the identity cannot be made for even a single non-zero $\varphi\in\DTd{d}$ then, when $\varphi=\psi\in\Lpd{2}{d}$ with unit norm, and the Schwartz bracket is replaced with the $\Lpd{2}{d}$ inner product then $E$ is a standard eigenvalue of the linear operator $H$ provided the potential makes sense as an operator of multiplication. Now since all $\varphi$ are also in the Hilbert space but in general an element of the Hilbert space is not a test function, this motivates the notion of a \textbf{proxy test function}.
\begin{defn}[Proxy Test Functions]
Proxy test functions are exactly all the elements of the operator domain being the closed dense subset $D(H)\subset\Lpd{2}{d}$ that make the usual notion of eigenvalue (elements of the point spectrum) correct with the Hilbert inner product replacing the Schwartz bracket in \ref{csp}. In addition, they include the set of generalized eigenvalues which are not elements of the Hilbert space, $\Lpd{2}{d}$.
\end{defn}
The justification is straightforward. Since the inner product will always entail the Lebesgue integral, the Schwartz bracket will also be given by the same expression with the proxy test functions, and moreover the acting linear functional for the Schwartz bracket will be exactly the Lebesgue integral. It is in precisely this notion in which the quantum theory is generalized. The C-spectrum extending the notion of eigenvalue, while the proxy test functions fill in the logical gap between the true test functions and that of the eigenspace for which the Hilbert inner product makes sense. Finally, recall that since the Hilbert space is the completion of the test function space with respect to the Hilbert inner product, there always exists a sequence of $\varphi_n\in\DTd{d}\fall n\in\N$ such that $\varphi_n \xrightarrow{n\rightarrow\infty}\psi\in\Lpd{2}{d}$, as well as $\varphi_n \xrightarrow{n\rightarrow\infty}\psi\in\DDd{d}$. Thus every proxy test function is given by a sequence of test functions, even when $\psi$ is not in the usual Hilbert space, i.e. when $\psi$ is a generalized eigenvector.

As for operators of multiplication, it is a well known theorem from functional analysis that every real valued function makes sense as an operator of multiplication and therefore is a self-adjoint operator on a suitable domain. Note that if the potential is $\Lpd{2}{d}$ the product still makes sense as an element of $\Lpd{1}{d}$, and thus the potential can vary from a typical function by as much as an arbitrary countably infinite set of Lebesgue measure zero. For the state of the art most general conditions on the potential, and a robust survey of the relevant techniques, see the monograph by \cite{JoLa}. Also, throughout this work, the binary equivalence relation $\,\dot{=}\,$ will be used to mean equality in the sense of distributions (only). It is now obvious explicitly that there is the need to work on the rigged Hilbert space or more aptly the Gelfand triple
\[
\DTd{d}\subset\Lpd{2}{d}\subset\DDd{d}.
\]

\section{Some Prerequisite Lemmas and Corollaries}

Throughout this work, the analysis is done in three Euclidean dimensions with its usual real topology. It is important to note the following interesting lemma with its corollary and its contrast subsequent lemma.

\begin{lem}[A Singular but Conditionally Trivial Distribution]
{
Let $(x_1,\,x_2,\,x_3) = x\in\R^3$ and $\delta(x)\in\DDd{3}$ such that for all $\varphi\in\DTd{3}$ one has $\left\langle\delta,\,\varphi\right\rangle = \delta[\varphi]=\varphi(0)$. Also, let $|x|^2 = x_1^2+x_2^2+x_3^2$ the Euclidean norm squared, then
\[
\left\langle\frac{\delta(x)}{|x|},\,\varphi\right\rangle = 0\quad \fall\varphi\in\DTd{3}
\]
}
\end{lem}
The proof is surprisingly straightforward and intuitively the result should be thought of roughly as so. Every positive even power of the Euclidean norm is necessarily a smooth function and products of smooth functions with test functions yield another test function. Multiplying by such a term would obviously give zero. However, one must show this abstractly for all test functions.
\begin{proof}
Note that in the sense of distributions for $x\in\R^d$,
\[
\frac{1}{2d}|x|^2\nabla^2\delta(x) \,\dot{=}\, \delta(x),
\]
and in the present case $d=3$. Then, it is immediate that
\[
\frac{1}{6}|x|\nabla^2\delta(x) \,\dot{=}\, \frac{\delta(x)}{|x|} \,\dot{=}\, S,
\]
where the singular distribution is named $S$ for brevity and ostensibly, $S\in\DDd{3}$. For any $\varphi\in\DTd{3}$, one has that
\begin{align}
\left\langle\frac{\delta(x)}{|x|},\,\varphi\right\rangle &= \langle\tfrac{1}{6}|x|\nabla^2\delta(x),\,\varphi\rangle\label{2.1} \\
&= \langle\nabla^2\delta(x),\,\tfrac{1}{6}|x|\varphi\rangle \\
&= \left\langle\delta(x),\,\nabla^2\left(\tfrac{1}{6}|x|\varphi\right)\right\rangle \\
&= \tfrac{1}{6}\left\langle\delta(x),\,\left(|x|\nabla^2\varphi + \tfrac{2x}{|x|}\cdot\nabla\varphi + 2\varphi\tfrac{1}{|x|}\right)\right\rangle \\
&= \tfrac{1}{3}\left\langle\delta(x),\,\tfrac{x}{|x|}\cdot\nabla\varphi\right\rangle + \tfrac{1}{3}\left\langle\delta(x),\, \tfrac{1}{|x|}\,\varphi\right\rangle \label{2.5}
\end{align}
Now the first term to the right of the equal sign in \eqref{2.5} must be zero due to continuity and linearity. In this instance, it requires that the order of component-wise evaluation of the linear functional is the same regardless of the order. For regard,
\begin{align*}
\left\langle\delta(x),\,\tfrac{x}{|x|}\cdot\nabla\varphi\right\rangle &= \left\langle\delta(x_1)\delta(x_2)\delta(x_3),\,\left(\frac{x_1}{\sqrt{x_1^2+x_2^2+x_3^2}}\varphi_{x_1} + \frac{x_2}{\sqrt{x_1^2+x_2^2+x_3^2}}\varphi_{x_2} + \frac{x_3}{\sqrt{x_1^2+x_2^2+x_3^2}}\varphi_{x_3}\right)\right\rangle \\
&= \left\langle\delta(x_1)\delta(x_2)\delta(x_3),\,\tfrac{x_1}{\sqrt{x_1^2+x_2^2+x_3^2}}\varphi_{x_1}\right\rangle + \left\langle\delta(x_1)\delta(x_2)\delta(x_3),\,\tfrac{x_2}{\sqrt{x_1^2+x_2^2+x_3^2}}\varphi_{x_2}\right\rangle \\
&\phantom{= \left\langle\delta(x_1)\delta(x_2)\delta(x_3),\,\tfrac{x_1}{\sqrt{x_1^2+x_2^2+x_3^2}}\varphi_{x_1}\right\rangle\:}+ \left\langle\delta(x_1)\delta(x_2)\delta(x_3),\,\tfrac{x_3}{\sqrt{x_1^2+x_2^2+x_3^2}}\varphi_{x_3}\right\rangle \\
&= \left\langle\delta(x_2)\delta(x_3),\,0\right\rangle + \left\langle\delta(x_1)\delta(x_3),\,0\right\rangle + \left\langle\delta(x_1)\delta(x_2),\,0\right\rangle \\
&= \left\langle\delta(x_1),\,\tfrac{x_1}{|x_1|}\varphi_{x_1}\right\rangle + \left\langle\delta(x_2),\,\tfrac{x_2}{|x_2|}\varphi_{x_2}\right\rangle + \left\langle\delta(x_3),\,\tfrac{x_3}{|x_3|}\varphi_{x_3}\right\rangle \\
&= 0.
\end{align*}
The third to last line is obviously zero. The second to last line is evidently zero owing to the fact that the signum function $\sgn(x_k)=0$ whenever $x_k=0$ for $k\in\{1,2,3\}$. This declaration for the signum function's behavior had to be checked against the requirements of linearity, continuity, and in this context the happenstance of the commutativity of the tensor product of delta functions in each coordinate component. It could not be assumed by fiat because the product of distributions is not necessarily another distribution.

The vanishing of this term now shows from \eqref{2.1} and the remaining term of \eqref{2.5} that,
\[
\left\langle\frac{\delta(x)}{|x|},\,\varphi\right\rangle = \frac{1}{3}\left\langle\frac{\delta(x)}{|x|},\,\varphi\right\rangle,
\]
which is valid for all $\varphi\in\DTd{3}$ and hence $S\,\dot{=}\, 0$, as claimed \qedhere
\end{proof}

This proof is generalized to all $d\in\N$ in a straightforward manner. It is worth noting however that for $d=1$ although $\tfrac{\delta}{|x|}\,\dot{=}\, 0$, there is also the result $\tfrac{\delta}{x}\,\dot{=}\,-\delta'$, where $\delta'$ is the derivative of the Dirac delta on $\R$. The correct way of viewing the $\sgn(x)$ function is as a distribution when it is on the left but as a piece-wise function when on the right. This is just another way of saying that the distributional derivative is not the same operation as the usual derivative. In particular, there is the very interesting result that the distributional derivative is continuous and hence as a linear operator, must be bounded. This is not possible for the usual derivative, except by contrivance via the Sobolev spaces. Through the weak topology, boundedness of the weak derivative comes for free, and in particular does not require the notion of a norm. In any case, it is useful to note the following immediate corollary.

\begin{cor}[Triviality for Smooth Functions]
Let $f\in C^\infty(\R^3)$ then
\[
\left\langle\frac{\delta}{|x|}f,\,\varphi\right\rangle = 0 \quad\fall\varphi\in\DTd{3}
\]
\end{cor}
\begin{proof}
Since for $f\in C^\infty(\R^3)$ the product $(f\cdot\varphi)\in\DTd{3}$. \qedhere
\end{proof}
Note also that the result does not require $d=3$ and holds for all $d\in\N$. The next lemma is in contrast to the previous two propositions.
\begin{lem}[Non-triviality]
Let $g\in\Ckd{k}{3}$ then
\begin{equation}
\left\langle\frac{\delta}{|x|}g,\,\varphi\right\rangle \neq 0 \quad\fall\varphi\in\DTd{3} \label{26}
\end{equation}
\end{lem}
\begin{proof}
The full proof is really just a logical consequence of the generic characterization theorem for distributions as a finite number of derivatives of a continuous function. But for simplicity and completeness simply consider $g(x)=|x|\in C^0(\R^3)$, then \eqref{26} returns simply $\varphi(0)$ which in general is not zero. This result again holds for all $d\in\N$.
\end{proof}

The central proposition has to do with a specific choice of the continuous but not smooth function $g$, as it pertains to the relevance of the solution to the system at hand.
\begin{prop}[ $(\delta\psi)\in\DDd{3}$ ]\label{21}
Let $x\in\R^3$ and choose $g(x) = e^{-b\,|x\,|}$ with $b\in\R$, then
\[
\left\langle\frac{\delta}{|x|}\,e^{-b\,|x\,|},\,\varphi\right\rangle = b\,\left\langle -\delta,\,\varphi\right\rangle
\]
\end{prop}
\begin{proof}
\begin{align*}
\left\langle\frac{\delta}{|x|}\,e^{-b\,|x\,|},\,\varphi\right\rangle &= \left\langle\frac{1}{6}|x|\nabla^2\delta(x)\,e^{-b\,|x\,|},\,\varphi\right\rangle \\
&= \left\langle\delta(x),\,\nabla^2\left(\frac{1}{6}|x|\,e^{-b\,|x\,|}\,\varphi\right)\right\rangle \\
&= \frac{1}{6}\left\langle\delta(x),\,\nabla\cdot\left(\tfrac{x}{|x|}\,e^{-b\,|x\,|}\,\varphi - b\,x\,e^{-b\,|x\,|}\,\varphi + |x|\,e^{-b\,|x\,|}\,\nabla\varphi\right)\right\rangle \\
&= \frac{1}{3}\left\langle\delta(x),\,\tfrac{1}{|x|}\,e^{-b\,|x\,|}\,\varphi\right\rangle - \frac{2}{3}\left\langle\delta(x),\, b\,e^{-b\,|x\,|}\,\varphi\right\rangle + \frac{1}{3}\left\langle\delta(x),\, e^{-b\,|x\,|}\,\tfrac{x}{|x|}\cdot\nabla\varphi\right\rangle \\
&\quad + \frac{1}{6}\left\langle\delta(x),\, b^2\,|x|\,e^{-b\,|x\,|}\,\varphi\right\rangle 
 - \frac{1}{3}\left\langle\delta(x),\, b\,e^{-b\,|x\,|}\,x\cdot\nabla\varphi\right\rangle  + \frac{1}{6}\left\langle\delta(x),\, |x|\,e^{-b\,|x\,|}\,\nabla^2\varphi\right\rangle \\
&= \frac{1}{3}\left\langle\delta(x),\,\tfrac{1}{|x|}\,e^{-b\,|x\,|}\,\varphi\right\rangle - \frac{2}{3}\left\langle\delta(x),\, b\,e^{-b\,|x\,|}\,\varphi\right\rangle \\
&= \frac{1}{3}\left\langle\frac{\delta}{|x|}\,e^{-b\,|x\,|},\,\varphi\right\rangle + \frac{2}{3}\, b\,\left\langle -\delta,\,\varphi\right\rangle
\end{align*}
Subtracting one third of the same expression on both sides of the last line gives the result as quoted in the proposition. \qedhere
\end{proof}

\section{The Singular Fundamental Solution}

So far only the theory of distributions has been needed and the theorems have simply been customized to a peculiar singular distribution. The remainder of the paper is dedicated to highly specific propositions which are needed to solve the formerly ill-posed spectral problem, now a well-posed $C$-spectral problem, where the Hamiltonian is given by the distribution,
\[
H\psi = -\frac{\hbar^2}{2m}\nabla^2\psi - \alpha\,\delta(x)\psi,
\]
and $\psi$ is determined by the $\Lpd{2}{3}$ solution of
\begin{equation}
-\frac{\hbar^2}{2m}\nabla^2\psi(x) = -|E|\psi(x). \label{31}
\end{equation}
The usual eigenvalue here $E$ is assumed to be $E<0$, but this is not a necessary assumption as it follows from the spectral properties of the Laplace operator on $\R^3$ that $E<0$ is a property of the real point spectrum for non-zero eigenfunctions.

The reader is invited at this point to learn on their own just how and why all standard methods of determining any usual definition of $E$ fail when written as,
\begin{equation}
H\psi = -\frac{\hbar^2}{2m}\nabla^2\psi - \alpha\,\delta(x)\psi = E\psi, \label{28}
\end{equation}
with $\hbar,\,m > 0$, $E < 0$, and $\alpha\in\R\setminus{\{0\}}$. To much surprise the above problem is resilient against the method of the Fourier transform for determining the resolvent, even though the Dirac delta is a tempered distribution, and is well behaved under the Fourier transform. The problem rather is that the $\Lpd{2}{3}$ Green's function or fundamental solution is singular at the point support of the Dirac measure. This problem turns out to be fatal. The author is aware of only two other methods for solving the system \eqref{28}. The first is the method of self-adjoint extensions of the Laplace operator as given in \cite{AGHH}. The second is by the Laplace transform. There is also a third possibility hinted in \cite{AGHH}, regarding nonstandard analysis. The status of which for \eqref{28} is unknown as of this writing.

Each method is mathematically rigorous but the only issue to be taken is the issue regarding physical but idealized experimental propositions. Specifying the wave function on a boundary (surface) is reasonable, but specifying it at a point (in all but $\R$) is unattainable because such a point or even countably infinite of them are a Lebesgue set of measure zero and cannot be known or even manipulated for a general wave function (element of $\Lpd{2}{d},\,d>1$). The consequence is that inserting initial conditions for the Laplace transform or inserting continuity or jump behavior may in fact be physically equivalent to pre-selecting the solution.

Instead, one solves \eqref{31} to start, which is well-posed on $\R^3$. It can be re-written as,
\begin{equation}
-\frac{\hbar^2}{2m}\nabla^2\psi + |E|\psi = 0. \label{33}
\end{equation}
The resolvent operator is found using the Fourier transform for the functional inverse relation, i.e. Green's function,
\[
\left(-\frac{\hbar^2}{2m}\nabla^2 + |E|\right)G = \delta.
\]
For purposes of application to the Hilbert space, the Fourier transform should be unitary. Additionally, the Fourier transform should transform the convolution identity above on the right of the equal sign (the Dirac delta) to the multiplicative identity 1, with no extraneous constants. The remaining issue is to decide the sign of the forward and backward Fourier transforms. Throughout this work, the kernel for the forward transform is chosen as $e^{-2\pi i k\cdot x}$ with $x,\,k\in\R^3$.

In any case, the solution for the above Green's function is
\begin{equation}
G(x) = \frac{m}{\pi\hbar^2}\frac{e^{-b\,|x|}}{|x|},
\end{equation}
with $b := \sqrt{2m|E|}/\hbar > 0 $. Therefore, the normalized (vector with unit norm) $\Lpd{2}{3}$ solution to \eqref{33} is
\begin{equation}
\psi(x) = \sqrt{\frac{b}{2\pi}}\frac{e^{-b\,|x|}}{|x|} \label{35}
\end{equation}
Notice that both $G(x)$ and $\psi(x)$ are not finite at $x=0$. But nonetheless, they are both $\Lpd{2}{3}$ as is required by the isometric automorphic Fourier map on the Hilbert space.

\section{The Singleton C-Spectral Set}

One must now construct the distribution $(H\psi)$ by applying the distributional Laplace operator to $\psi$ and placing the wave function next to the pseudo-potential as so,
\begin{align}
H\psi &\,\dot{=}\, -\frac{\hbar^2}{2m}\nabla^2\psi(x) - \alpha\delta(x)\psi(x) \nonumber \\
&\,\dot{=}\, -\frac{\hbar^2}{2m} \sqrt{\frac{b}{2\pi}} \nabla^2\left(\frac{e^{-b\,|x|}}{|x|}\right) - \alpha \,\sqrt{\frac{b}{2\pi}} \delta(x)\frac{\,e^{-b\,|x|}}{|x|} \nonumber \\
&\,\dot{=}\, -\frac{\hbar^2}{2m} \sqrt{\frac{b}{2\pi}}\left(-4\pi\delta(x) + b^2\,\frac{\,e^{-b\,|x|}}{|x|}\right) - \alpha \,\sqrt{\frac{b}{2\pi}} \delta(x)\frac{\,e^{-b\,|x|}}{|x|} \nonumber \\
&\,\dot{=}\, \sqrt{\frac{b}{2\pi}}\left(\frac{4\pi\hbar^2}{2m}\delta -\alpha\,\frac{\,\delta}{|x|}\,e^{-b\,|x|} \right) - \frac{\hbar^2}{2m}b^2\,\psi \nonumber \\
&\,\dot{=}\, \sqrt{\frac{b}{2\pi}}\left(\frac{4\pi\hbar^2}{2m}\delta -\alpha\,\frac{\,\delta}{|x|}\,e^{-b\,|x|} \right) + E\,\psi  \label{41}
\end{align}
Based on the results of proposition \ref{21}, it is clear that $(H\psi)\in\DDd{3}$. Thus, it now makes sense to solve the well-posed expression,
\[
\left\langle H\psi,\,\varphi\right\rangle = E \,\left\langle\psi,\,\varphi\right\rangle
\]
The above expression when combined with \eqref{41} leads to
\[
\sqrt{\frac{b}{2\pi}}\left\langle\left(\frac{4\pi\hbar^2}{2m}\delta -\alpha\,\frac{\,\delta}{|x|}\,e^{-b\,|x|} \right),\,\varphi\right\rangle + E\,\left\langle\psi,\,\varphi\right\rangle = E\,\left\langle\psi,\,\varphi\right\rangle
\]
With the only energy eigenvalue, $E$, term on the right hand side canceling with an identical term on the left, the remaining statement is,
\begin{align*}
\left\langle\left(\frac{4\pi\hbar^2}{2m}\delta -\alpha\,\frac{\,\delta}{|x|}\,e^{-b\,|x|} \right),\,\varphi\right\rangle = \left\langle\left(\frac{4\pi\hbar^2}{2m}\delta  + \alpha\,b\,\delta\right),\,\varphi\right\rangle &= 0 \\
&= \left\langle\left(\frac{4\pi\hbar^2}{2m} + \alpha\,b\right)\,\delta,\,\varphi\right\rangle = 0.
\end{align*}
The first term's equality on the top far left uses the results of proposition \ref{21}, while the bracket in the bottom right shows that the expression in parenthesis must vanish for all $\varphi\in\DTd{3}$. Recalling that $\hbar,\,m,\,b > 0$ and $\alpha\in\R\setminus{\{0\}}$, this leads to the requirement that
\[
\frac{4\pi\hbar^2}{2m} + \alpha\,b = 0,
\]
and consequently,
\begin{equation}
E = - \frac{2\pi^2\hbar^6}{m^3\,\alpha^2} \label{bnden}
\end{equation}
This demonstrates that the presence of the Dirac delta function as a potential in the Schrodinger equation concentrates or "crystallizes" the point spectrum of the free Hamiltonian. The $\Lpd{2}{3}$ condition for the free Hamiltonian (the Laplace operator) is that $E<0$, which forces $b>0$, and thus allows $\psi$ to be in the Hilbert space. Mathematically, it is stated that the free Hamiltonian operator, $H_0$, has the point spectral property,
\[
\sigma_\text{p}(H_0) = (-\infty,\,0)\subset\R\subset\C.
\]
Whereas it has just been proved from above that,
\begin{thm}[The C-spectrum of $(H\psi)\in\DDd{3}$ from \eqref{28} ]
Given $m,\,\hbar > 0$ and $\alpha\in\R\setminus\{0\}$, the C-spectrum of the system's formal Hamiltonian, $H$ from \eqref{28}, $x\in\R^3$ with the Dirac delta measure as a pseudo-potential, is given by the singleton set,
\[
\sigma_C(H) = \left\{- \frac{2\pi^2\hbar^6}{m^3\,\alpha^2}\right\}\subset\R\subset\C,
\]
with,
\[
0 > - \frac{2\pi^2\hbar^6}{m^3\,\alpha^2} \in \R.
\]
\end{thm}
There is an immediate corollary that connects this result to the literature.
\begin{cor}[Correspondence with Self-Adjoint Extensions]
When $\hbar^2/m = 2$, then the C-spectrum singleton set becomes
\[
\sigma_C(H) = \left\{- \frac{16\pi^2}{\alpha^2}\right\} = \{E_0\},
\]
where $E_0$ is exactly the single simple eigenvalue on p. 29 of \cite{AGHH} part (b) of theorem 1.3.1. Notice that the coupling constant, $\alpha$ in \cite{AGHH} for this SPP is upside down, as is evident from the domain of definition stated in relation 1.1.12 on p. 13 of \cite{AGHH}. Thus, the two results coincide precisely.
\end{cor}
\begin{proof}
The proof is elementary. Simply set $\tfrac{\hbar^2}{m} = 2$ to conclude that $\tfrac{\hbar^6}{m^3} = 8$, while at the same time when comparing with \cite{AGHH} note that (using the symbols, $E_0$, \text{\boldmath$\alpha$} and $k_0$ as defined there)
\[
E_0 = k_0^2 = (-4\pi\,i\,\text{\boldmath$\alpha$})^2 = -16\pi^2\text{\boldmath$\alpha$}^2 = \frac{-16\pi^2}{\alpha^2},
\]
where $\text{\boldmath$\alpha$}$ is the (multiplicative inverse of the) coupling constant $\alpha$ from \ref{28} used in this work. \qedhere
\end{proof}
Finally, one has a rigorous distributional version of the Hellmann-Feynman theorem,
\begin{thm}[Hellmann-Feynman for Distributions]
\[
\left\langle\frac{\partial}{\partial b}\frac{\delta}{|x|}\,e^{-b\,|x\,|},\,\varphi\right\rangle = \frac{\partial}{\partial b}\;\left\langle -b\,\delta,\,\varphi\right\rangle
\]
\end{thm}
\begin{proof}
From the results of proposition \ref{21}, the validity of the expression for all $b\in\R$, the smoothness of the exponential function on the left, the continuity of the Schwartz linear functional, and the smoothness of the monomial in $b$ on the right, the above quoted result follows.
\begin{align*}
\left\langle\frac{\partial}{\partial b}\frac{\delta}{|x|}\,e^{-b\,|x\,|},\,\varphi\right\rangle &= \frac{\partial}{\partial b}\;\left\langle -b\,\delta,\,\varphi\right\rangle \\
\left\langle -\,\delta\,e^{-b\,|x\,|},\,\varphi\right\rangle &= -\left\langle\delta,\,\varphi\right\rangle \qedhere
\end{align*}
This result holds for all $d\in\N$, as well.
\end{proof}

\section{Acknowledgments}
The author would like to thank the Burton Jones Chair of Pure Mathematics, Distinguished Professor Michel L. Lapidus for his helpful comments and corrections, as well as Alina Pineiro Escalera for her suggestions and commentary on the article, and finally Professor of Mathematics Marat Markin for his keen eye spotting a typo.

\end{document}